# The origin of hydrogen around HD 209458b

A. Lecavelier des Etangs[1,2], A. Vidal-Madjar[1,2] & J.-M. Désert[1,2]

[1]CNRS, UMR 7095, Institut d'Astrophysique de Paris, 98bis boulevard Arago, F-75014 Paris, France.
[2]UPMC Université Paris 6, UMR 7095, IAP, 98bis boulevard Arago, F-75014 Paris, France.

e-mail: lecaveli@iap.fr

Using numerical simulation, Holmström *et al.*[1] proposed a plausible alternative explanation of the observed Lyman-$\alpha$ absorption that was seen during the transit of HD 209458b (ref. 2). They conclude that radiation pressure alone cannot explain the observations and that a peculiar stellar wind is needed. Here we show that radiation pressure alone can in fact produce the observed high-velocity hydrogen atoms. We also emphasize that even if the stellar wind is responsible for the observed hydrogen, to have a sufficient number of atoms for charge exchange with stellar wind, the energetic neutral atom (ENA) model also needs a significant escape from the planet atmosphere of similar amplitude as quoted in ref. 2.

The simulation of ref. 1 is aimed at reproducing the observed absorption spectrum in Lyman-$\alpha$ with 15 ± 4% absorption between −130 and 100 km s$^{-1}$ (refs 2, 3). A mechanism is needed to produce hydrogen atoms at these high velocities exceeding the planet escape velocity. We previously proposed that hydrogen atoms in the exosphere are naturally accelerated by the stellar radiation pressure[2,4]; however, Holmström *et al.*[1] concluded that radiation pressure alone cannot explain the observation. Nonetheless, in their work, the strength of the radiation pressure has been artificially reduced to a value 2 to 5 times lower than the solar value, whereas the observed Lyman-$\alpha$ line strength and profile shows that it is significantly larger than the solar value. The low radiation pressure assumed by Holmström *et al.*[1] is valid only at high radial velocity. However, if low radiation pressure is assumed, high velocities are not reached, which therefore explains the different conclusion reached by Holmström *et al.* A correct treatment of the link between the radiation pressure and radial velocity is needed.

To show that the radiation pressure can explain the observed spectrum, we calculated the modelled Lyman-α profile with radiation pressure alone, in the same way as done in Fig. 3 of ref. 1 for the ENA model. This calculation is done taking into account the strength and profile of the Lyman-α line, and the corresponding variation of radiation pressure as a function of radial velocity. Planetary and stellar gravities are also included. These differences explain the different results obtained with radiation pressure alone in the two models. The result plotted in Fig. 1 shows that the resulting profiles are similar in the two models (radiation pressure alone and ENA with reduced radiation pressure), and neither possible model can be favoured. Radiation pressure cannot be excluded as an explanation of the observed spectrum.

Although we agree that the ENA model is a plausible scenario, we do not believe that ENAs can explain the observations better than a classical scenario with radiation pressure. The ENA model requires extraordinary conditions for the wind parameters (high temperature and low velocity) which are not constrained by any other observations, whereas the radiation pressure as measured in the Lyman-α spectrum can self-consistently explain the observations.

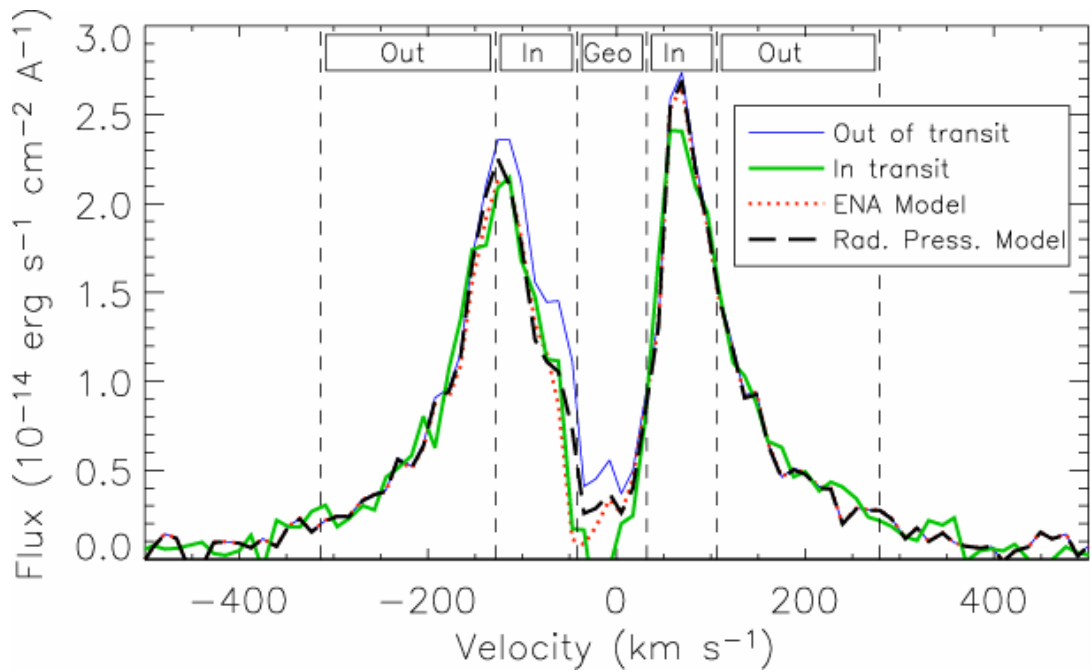

**Figure 1 Comparison of modelled with observed Lyman-α profiles as in Fig. 3 of ref. 1**. The thin blue line and the thick green line are for the observed profile before and during the transit, respectively. The red dotted line is for the modelled profile from the ENA model with reduced radiation pressure. The black dashed line is for the modelled profile computed using a model with radiation pressure. This last profile fits the data well with a $\chi^2$ of 45 for 40 degrees of freedom. As the profile for the radiation pressure model is similar to the one for the ENA model, neither possible model can be favoured. Radiation pressure cannot be excluded as an explanation for the observed spectrum.